# Superconducting Order from Disorder in 2H-TaSe$_{2-x}$S$_x$

Lijun Li[1,2,†], Xiaoyu Deng[3], Zhen Wang[1], Yu Liu[1], A. M. Milinda Abeykoon[4], E. Dooryhee[4], A. Tomic[5], Yanan Huang[1,*], J. B. Warren[6], E. S. Bozin[1], S. J. L. Billinge[1,5], Y. P. Sun[2,7], Yimei Zhu[1], G. Kotliar[1,3] and C. Petrovic[1,†]

[1]Condensed Matter Physics and Materials Science Department, Brookhaven National Laboratory, Upton, New York 11973, USA
[2]Key Laboratory of Materials Physics, Institute of Solid State Physics Chinese Academy of Sciences Hefei 230031 China
[3]Department of Physics & Astronomy, Rutgers, The State University of New Jersey - Piscataway, NJ 08854, USA
[4]Photon Sciences Directorate, Brookhaven National Laboratory, Upton NY 11973 USA
[5]Department of Applied Physics and Applied Mathematics, Columbia University, New York NY 10027 USA
[6]Instrumentation Division, Brookhaven National Laboratory, Upton, New York 11973, USA7
[7]High Magnetic Field Laboratory, Chinese Academy of Sciences, Hefei 230031 China

**We report on the emergence of robust superconducting order in single crystal alloys of TaSe$_{2-x}$S$_x$ (0 ≤ x ≤ 2) . The critical temperature of the alloy is surprisingly higher than that of the two end compounds TaSe$_2$ and TaS$_2$. The evolution of superconducting critical temperature T$_c$ (x) correlates with the full width at half maximum of the Bragg peaks and with the linear term of the high temperature resistivity. The conductivity of the crystals near the middle of the alloy series is higher or similar than that of either one of the end members 2H-TaSe$_2$ and/or 2H-TaS$_2$. It is known that in these materials superconductivity (SC) is in close competition with charge density wave (CDW) order. We interpret our experimental findings in a picture where disorder tilts this balance in favor of superconductivity by destroying the CDW order.**

## Introduction

The interplay of disorder and interactions is a fruitful area of investigation. In the absence of electron-electron interactions, disorder can turn a metallic system into an Anderson insulator (1), but can remain metallic when interactions are important. The additional complexity of competing orders such superconductivity with charge density wave (CDW) or magnetism makes this problem one of the most challenging frontiers in physics (2-5). A large body of literature is devoted to this interplay in nearly magnetic materials (6). The interplay of disorder and superconductivity in CDW materials has been less explored than its magnetic analog.

Superconductivity and CDW are traditionally viewed as weak-coupling Fermi surface instabilities due to electron-phonon coupling (7). Arguments have been made both for their cooperation and competition (8, 9). Hexagonal transition metal dichalcogenide 2H-TaSe$_2$ (*P63/mmc* space group) undergoes a second-order transition to an incommensurate CDW at 122 K followed by a first-order lock-in transition to a commensurate CDW (CCDW) phase at 90 K, eventually becoming superconducting below 0.14 K upon cooling (10, 11). 2H-TaS$_2$ has Tc = 0.8 K below an in-plane CCDW at 78 K (10, 12). The CDW mechanism in 2H-TaSe$_2$ involves an electron instability in the bands nested away from the Fermi surface, whereas 2H-TaS$_2$ features a polar charge and orbital order (13, 14). CDW in 2H-TaSe$_2$ is dominated by hopping between next-nearest neighbors that creates three weakly coupled triangular sublattices (15). It is of interest to note that the 2H-TaSe$_2$ is quasi-two dimensional (2D) metal with pseudogap and with c-axis resistivity 25-50 times higher than the in-plane resistivity, i.e. $\rho_c(T) \gg \rho_{ab}(T)$ (10,13,16-18).

Here we report that in the 2H-TaSe$_{2-x}$S$_x$ alloy series the CDW is suppressed and the superconductivity is maximized with crystallographic disorder. The T$_c$(x) evolution is correlated with the high temperature linear resistivity $\rho$(T)=aT+b. The constant term b can be attributed to impurity-like carrier scattering off the local CDW fluctuations, and it also appears in dynamical mean field theory (DMFT) of bad metals at high temperature (19-22). On very general grounds (Anderson theorem) *s*-wave superconductivity is immune to weak disorder (23), on the other hand disorder is detrimental to CDW. We argue that the increase in superconducting T$_c$ in the alloy is a direct result of disorder induced suppression of CDW order. In a weak coupling picture, CDW suppression results in an increase in number of carriers available for superconductivity pairing at Fermi surface, thus enhancing T$_c$. The physical scenario that in systems where CDW competes with superconductivity disorder promotes the latter is very general and extends to a strongly coupled situation as long as disorder remains weak (Supplement).

**Results**

Powder patterns for all samples have been successfully indexed within the *P63/mmc* space group. Representative refinement is shown in Fig. 1(a) and the unit cell is shown in Fig. 1(b). Single crystal x-ray diffraction patterns for a subset of single crystals used in this study [Fig. 1(c)] show (00l) reflections. Reflections shift to higher scattering angles with increasing S indicating decrease of the unit cell volume. Evolution of unit cell parameters with S, obtained from fits to the powder patterns [Fig. 1(d) and (e)], is consistent with the single crystal data.

The resistivity of all single crystals [Fig. 2(a)] is metallic. The curves for $0 \leq x \leq 0.25$ show a change of slope in $\rho(T)/\rho(200K)$ [Fig. 2(b)]. As opposed to commonly observed increase in $\rho(T)$ at $T_{CDW}$ the slope change is attributed to CDW transition that leaves the bands associated with the undistorted sublattice ungapped (14-17). The hump shifts to lower temperature with S doping and vanishes for $x \geq 0.52$, but appears again for $x = 1.98$ at about 70 K, somewhat below the $T_{CDW} = 75$ K for pure 2H-TaS$_2$ [Fig. 2(c)]. The resistivity decreases to zero at lower temperatures, implying superconductivity [Fig. 2(d)]. The magnetic susceptibility transitions and the large values of $-4\pi\chi$ at 1.9 K imply bulk superconductivity [Fig. 3(a)].

Specifically, the anisotropic M(H) curves [Fig. 3(b)] confirm type-2 superconductivity for TaSe$_{1.48}$S$_{0.52}$ and imply anisotropic critical current density. The lambda anomaly in the specific heat jump around T = 4 K [Fig. 3(c)] is suppressed significantly in 4 T. A rough estimate of the average electron phonon coupling $\lambda_{e-ph}\sim$ 0.73(1) can be obtained from the McMillan equation assuming the empirical value of the Coulomb pseudopotential $\mu^*=0.15$ and taking the Debye frequency as the relevant phonon energy (24):

$$\lambda = \frac{\mu^* \ln\left(\frac{1.45 T_C}{\theta_D}\right) - 1.04}{1.04 + \ln\left(\frac{1.45 T_C}{\theta_D}\right)\left(1 - 0.62 \mu^*\right)} \qquad (1)$$

When compared to the parent 2H-TaSe$_2$ with electronic specific heat coefficient $\gamma$ = 4.5 mJ mol$^{-1}$ K$^{-2}$, $\gamma$ is larger for 50% S-doped sample [Fig. 3(d), Table I]. The electron-phonon coupling $\lambda_{e-ph}$= 0.73 is somewhat larger than for 2H-TaSe$_2$ and 2H-TaS$_2$ (Table I).

The ratio of the gap at the critical temperature $2\Delta/k_B T_c = 2.17$ can be obtained by linear fitting $\ln(C_e/\gamma T_c)$-$T_c/T$ data [Fig. 3(d) inset].

In a multiband electronic system with local CDW fluctuations (15-17,20) above $T_{CDW}$, such as 2H-Ta(Se,S)$_2$, the carrier scattering mechanism arises from collective excitations below the CDW and from local CDW fluctuations above the CDW (19). Above $T_{CDW}$ $\rho(T) \sim aT+b$, immediately below $T_{CDW}$ $\rho(T) \sim dT^2$ and at temperatures below about 15 K -20 K $\rho(T) \sim cT^5$. The $T^5$ is due to normal electron-phonon scattering whereas the $T^2$ arises due to scattering of electrons by collective excitations of CDW; the rapid drop just below CDW is due to CDW phase ordering. The linear terms a, b above $T_{CDW}$ arise due to electron-phonon scattering and phase disorder impurity-like scattering due to local CDW fluctuations. The fits of resistivity for the entire single crystal alloy series of 2H-TaSe$_{2-x}$S$_x$ (0 ≤ x ≤ 2) are satisfactory [Fig. 4(a-c), Table II]. In all samples resistivity is linear at high temperatures. Just above the superconducting $T_c$ interband scattering is negligible and individual *s*- and *d*-band normal electron phonon scattering dominates (19). The constant term b for such crystals is much smaller than for CDW samples, because disorder suppresses the CDW therefore increases the number of carriers and thus the conductivity.

We present the evolution of superconducting $T_c$ in 2H-TaSe$_2$ with S substitution x (normalized to $T_c$ value for x = 0) in Fig. 4(d). The $T_c$ shows 30-fold increase and anticorrelation with the evolution of the high temperature local charge fluctuation parameter b(x) (also normalized to b value for x=0) [Fig. 4(d)]. Note that weak decrease of $T_c$ near x = 1 coincides with weak increase in normalized b(x) near the same S content. It appears that the considerable increase and evolution of $T_c(x)$ is related to an increase in available carrier concentration or mobility. These changes are matched [Fig. 4(d)] with the nearly identical evolution of the crystallographic disorder as shown by the width of diffraction peaks taken on single crystals (FWHM) normalized to width of x=0 crystal. The linear resistivity aT+b in the high-$T_c$ crystals without CDW is rather close to the Bloch-Grüneisen phonon resistivity (Fig. 4(c), Supplement), suggesting that the strong suppression of parameter b(x) in 2H-TaSe$_{2-x}$S$_x$ can be attributed to suppression of local CDW fluctuations for high-$T_c$ crystals.

Figure 4(e) presents the electronic phase diagram. With the increase in x, the CDW transition of 2H-TaSe$_2$ is suppressed, whereas the superconducting transition temperature $T_c$ increases up to 4.2 K for x = 0.52 where CDW disappears. Further sulfur increase shows weak but well resolved minimum in $T_c$(x) for x = 1.10 up to the second maximal value of 4.28 K for x = 1.65 sulfur content in 2H-TaSe$_{1-x}$S$_x$. Signature of a CDW state, most probably CDW of the pure 2H-TaS$_2$, appears in $\rho$(T) for higher S content up to x=2. The two $T_c$ maxima in the double-dome appear at the critical doping where CDW orders vanish similar to 1T-Se$_2$ and 2H-NbSe$_2$ (25,26).

**Discussion**

It should be noted that electron-irradiated 2H-TaSe$_2$ shows enhancement of superconducting $T_c$ up to about 2.5 K (27). Irradiation introduces defects, i.e. changes in stoichometry similar to chemical substitutions. Cu-intercalated 2H-TaS$_2$ shows enhancement of $T_c$ up to 4.7 K (28). Copper behaves as n-type dopant and therefore its intercalation brings both disorder and charge transfer (29). Similar doping and disorder interplay is expected with Na intercalation in 2H-TaS$_2$ where $T_c$ was raised up to 4.4 K (30). In contrast, isoelectronic substitution in 2H-TaSe$_{2-x}$S$_x$ single crystal alloy series allows for clear separation of disorder from doping-induced changes. The basic electronic structure of 2H-TaSe$_2$, 2H-TaSeS and 2H-TaS$_2$ is quite similar as shown in the LDA calculations (Supplementary material), leaving disorder as the origin of the increase in T$_c$. The intense variation of physical properties at low energy is a prime example of emergent phenomena.

Below the CDW transition temperature a gap opens in 2 H-TaSe$_2$ on two distorted sublattices in contrast to undistorted sublattice (15). This makes for a small density of states at the Fermi level and consequently low superconducting T$_c$ (10). Sulfur substitution could introduce different Ta-S and Ta-Se bond lengths, disorder and puckering of metal plane. This would suppress CDW and increase density of electronic states, electron-phonon coupling and superconducting T$_c$ (31). The lattice defect disorder increases superconducting T$_c$ in CDW superconductor ZrTe$_3$ by factor of only 2-3 (32). Furthermore, T$_c$(x) [Fig. 4(e)] cannot be explained by the $f$-wave model of CDW that predicts linear T$_c$(x) in 2H-TaSe$_{2-x}$S$_x$ (0 ≤ x ≤ 2) whereas increased conductivity in high-T$_c$ crystals argues against the change of the amplitude of ionic vibrations as in disordered films or amorphous lattices (33,34).

The normal state properties of this material are not well understood theoretically. Our experimental results indicate that the low temperature specific heat coefficient of the alloy is very close to that computed in LDA (please see Supplementary Material), suggesting that the correlations due to Coulomb interactions are weak, while the electron phonon coupling couples strongly to a few states not too close to the Fermi surface, which is consistent with the results of ref 35. Alternatively, strong correlations are invoked in the exciton liquid model of 2H-TaSe$_2$. This model of CDW in 2H-TaSe$_2$ provides an explanation of some anomalous normal state properties such as linear resistivity above CDW transition, pseudogap, optical conductivity $\sigma(\omega,T)$ and incoherent metal features (20,36). Within that model and in contrast to $ZrTe_3$ (32), emergence of CDW reduces incoherent scattering, i.e. CDW-related bump in $\rho(T)$ is a coherence restoring transition that enables higher conductivity below $T_{CDW}$. The reduction in interband mixing dispersion that mixes the small number of $d_{z2}$ electrons and $p_z$ holes should remove the CDW, maximizing strong scattering off preformed incoherent excitons and enabling linear resistivity to progress to lower temperatures (20). The observation of wide temperature range of linear resistivity in high-$T_c$ crystals [Fig. 2(a)] suggests the rapid reduction in conduction and valence band mixing within that model. This calls for photoemission studies our newly synthesized alloy to test these theoretical models and to clarify the measure of the correlation strength.

Pressure should bring phonon hardening following the contraction of lattice parameters from $2H$-$TaSe_2$ to $2H$-$TaS_2$ (Fig. 1) (23,37,38). Assuming similar bulk modulus to $WSe_2$ (72 GPa) (39), the estimated chemical pressure differences of $2H$-$TaSeS$ ($T_c = 3.7$ K) when compared to $2H$-$TaSe_2$ ($T_c = 0.14$ K) and $2H$-$TaS_2$ ($T_c = 0.8$ K) are 6 GPa (positive pressure/contraction) and 2.8 GPa (negative pressure/expansion). Positive pressure increases considerably superconducting $T_c$ in both $2H$-$TaS_2$ and $2H$-$TaS_2$; 6 GPa brings $T_c$ in $2H$-$TaSe_2$ up to about 3 K (40). CDW is robust, surviving up to 20 GPa ($2H$-$TaSe_2$) and up to 16 GPa ($2H$-$TaS_2$). Clearly, chemical pressure may influence the rise of superconducting $T_c$ in S-doped $2H$-$TaSe_2$, but it cannot explain the absence of CDW in high-$T_c$ samples in the phase diagram [Fig. 4(e)], the increase of superconducting $T_c$ with lattice expansion in $2H$-$TaS_2$ or the $T_c(x)$ evolution in $2H$-$TaSe_{2-x}S_x$ ($0 \leq x \leq 2$) [Fig. 4(e)]. The reduction in conduction and valence band mixing within DMFT framework facilitates not only the reduction of incoherence and stabilization of pseudogap characterized by linear $\rho(T)$, but also an increase in density of states at the Fermi level thus highlighting the effect of disorder, incoherent states and the importance of local dynamical correlations (20).

In summary, we show that disorder-induced superconducting states arise by isoelectronic substitution in 2H-TaSe$_2$. In contrast to all known charge-density-wave CDW superconductors that have hitherto featured only a single dome of $T_c$ with variation of any external parameter, the electronic phase diagram we present features a weak double dome in $T_c(x)$. The increase in superconducting $T_c$ and changes in $T_c(x)$ are directly correlated with crystallographic disorder and disorder-induced scattering off the local CDW fluctuations. Our experimental findings can be understood on more general grounds without relying on a specific microscopic theory. For a given band structure, weak disorder does not affect the superconductivity of an *s*-wave superconductor (Anderson's theorem, Supplement) (1,23), but it is detrimental to the competing CDW order. The combination of these effects, results in an enhanced superconducting critical temperature and a reduction of the CDW.

## Materials and methods

Single crystals of 2H-TaSe$_{2-x}$S$_x$ ($0 \leq x \leq 2$) were grown via iodine vapor transport method. The source and growth zone were set at $900^\circ$C for 3 days and then kept at $900^\circ$C and $800^\circ$C, respectively, for 10 days. Black plate- like single crystals with a typical size of $3 \cdot 3 \cdot 0.2$ mm$^3$ were obtained. The element analysis was performed using an energy-dispersive x-ray spectroscopy (EDX) in a JEOL LSM-6500 scanning electron microscope. Electrical resistivity, specific heat and magnetization measurements were performed in a Quantum Design PPMS-9 and MPMS XL-5. X-ray diffraction (XRD) patterns on single crystals were taken using a Rigaku Miniflex. Room temperature powder XRD measurements were carried out at the X-ray Powder Diffraction (XPD, 28-ID-C) beam line at National Synchrotron Light Source II. The raw room temperature powder X ray 2D data were integrated and converted to intensity versus scattering angle using the software Fit2D (41). The average structure was assessed from raw diffraction data using the General Structure Analysis System (GSAS) operated under EXPGUI utilizing *P63/mmc* model from the literature (42-45).

**Acknowledgements:** Work at Brookhaven is supported by the U.S. DOE under Contract No. DE-SC00112704. Work at Institute of Solid State Physics of CAS is supported by the National Natural Science Foundation of China, Grant No. 11404342. Use of the National Synchrotron Light Source II, Brookhaven National Laboratory, was supported by the U.S. Department of Energy, Office of Science, Office of Basic Energy Sciences, under Contract No. DE-SC0012704. X. Deng  is supported by AFOSR MURI program. Y. L. and G. Kotliar are supported by U.S. Department of energy, Office of Science, Basic Energy Sciences as a part of the Computational Materials Science Program.

**Additional Information:** †To whom correspondence may be addressed. E-mail: petrovic@bnl.gov and lilijun@issp.ac.cn; *Present address: Key Laboratory of Materials Physics, Institute of Solid State Physics, Chinese Academy of Sciences, Hefei 230031, China.

**Author Contributions:** C. P. designed research. L. L. and Y. L. made crystals and carried out transport, magnetization and thermal measurements.  L. L. and Y. P. S. contributed single crystal X ray diffraction data. Y. L., Y. H. and J. W performed SEM measurements. Z. W. and

Y. Z. performed TEM measurements. A. M. M. A., E. D, A. T., E. S. B and S. J. L. B. carried out and analyzed crystal powder X ray diffraction data. C.P. supervised the project, analyzed the transport data with Y. L. and wrote the paper with L. L and with contributions from G. K. and Y. P. S.. X. D performed the LDA calculations and contributed to the theoretical interpretation of results. The manuscript reflects contribution and ideas of all authors.

**Competing Interests:** The authors declare no competing interests.

**FIGURE LEGENDS**

**Figure 1 Crystal Structure Aspects of 2H-TaSe$_{2-x}$S$_x$ (0≤x≤2)**

(a) Powder x-ray diffraction pattern for 2H-TaSe$_2$ at 300 K, shown as scattering intensity versus momentum transfer Q, indexed within *P63/mmc* space group. Crosses are data, solid red line is the model, green solid line is the difference (offset for clarity), and vertical ticks mark are the reflections.

(b) Structural motif of the *P63/mmc* model. Red dashed box depicts the unit cell.

(c) Single crystal x-ray diffraction patterns at the room temperature. Patterns are offset for clarity.

(d,e) Room temperature evolution of a and c lattice parameters, respectively, as obtained from powder diffraction data. Shaded is the range where CDW cannot be detected in resistivity.

**Figure 2 Electrical resistivity of 2H-TaSe$_{2-x}$S$_x$ (0≤x≤2)**

(a) Temperature dependence of the resistivity for single crystals in the absence of magnetic field.

(b,c) Temperature dependence of $\rho(T)/\rho(200\ K)$ near CDW transitions.

(d) $\rho(T)$ curves near superconducting transitions, indicating large enhancement of superconductivity.

**Figure 3 Magnetic and thermodynamic properties of 2H-TaSe$_{2-x}$S$_x$ (0≤x≤2)**

(a) Magnetic susceptibility after zero-field-cooling (ZFC, filled) and field-cooling (FC, open symbols). The smaller magnetization value for FC is likely due to the complex magnetic flux pinning effects.

(b) Magnetization hysteresis loops M(H) of TaSe$_{1.48}$S$_{0.52}$ for H ∥ ab (solid) and H ∥ c (open symbols).

(c) Low temperature specific heat of TaSe$_{1.48}$S$_{0.52}$ measured at H = 0 (solid) and in 4 T (open symbols).

(d) The electronic specific heat in the superconducting state C$_e$ for 2H-TaSe$_{1.48}$S$_{0.52}$ is obtained by subtracting the lattice contribution from the total specific heat: C$_e$ = C−C$_{ph}$ = γT where C$_{ph}$(T)= βT$^3$+δT$^5$ and θ$_D$=[(n·1.944·10$^6$)/β]$^{1/3}$ where n is the number of elements per formula unit. Inset: below the superconducting transition temperature, electronic specific heat temperature dependence follows an exponential decay, as C$_e$ ∼ exp[−Δ(T )/k$_B$T]. The solid line shows C$_e$/T calculated by assuming an isotropic *s*-wave BCS gap with 2Δ=k$_B$T$_c$ = 2.17.

### Figure 4 Electronic phase diagram of 2H-TaSe$_{2-x}$S$_x$ (0≤x≤2)

(a-c) Electronic scattering mechanism for (a) 2H-TaSe$_2$, (b) 2H-TaS$_2$ and (c) 2H-TaSe$_{0.9}$S$_{1.1.}$ Low temperature phonon scattering, scattering off collective excitations of CDW and high temperature scattering off the local CDW fluctuations are shown by blue, green and red solid lines respectively. Dashed violet line shows the phonon resistivity approximated by the Bloch-Grüneisen formula (see text and supplement). (d) Note that the weak double dome evolution of superconducting T$_c$(x) coincides with similar evolution of crystallographic disorder with sulfur content as revealed by full width at half maximum of [006] Bragg peak in Fig. 1(c). The intensity was normalized to 1 for each value of x. Moreover, T$_c$(x) is in close correlation with disorder-induced changes in the high-temperature local charge fluctuations as seen by the changes in parameter b; all normalized to values of 2H-TaSe$_2$ (x=0). The parameter b is shown from fits without (open) and with phonon subtraction from Bloch-Grüneisen formula (full symbols) (Supplement).

The error bars are about 0.01 for x, up to 0.04 for normalized $T_c$ and up to 0.01 for the

normalized parameter b and FWHM (also see Table 1). (e) Phase diagram indicating the

evolution of charge density wave (incommensurate ICCDW and commensurate CCDW) and

SC states with the change of x.

**Tables**

**Table I.** Superconducting parameters of 2H-TaSe$_2$, 2H-TaSe$_{1.48}$S$_{0.52}$ and 2H-TaS$_2$.

| Parameters | $2H$-TaSe$_2$ | $2H$-TaSe$_{1.48}$S$_{0.52}$ | $2H$-TaS$_2$ |
|---|---|---|---|
| $T_c$(K) | 0.14 | 4.20(1) | 0.8 |
| $\gamma$ (mJ/moleK$^2$) | 4.5 | 12.0(3) | 7.5 |
| $\beta$(mJ/moleK$^4$) | 0.72 | 0.65(2) | 0.44 |
| $\delta$(mJ/moleK$^6$) | - | 6(1)·10$^{-4}$ | - |
| $\lambda_{e\text{-}ph}$ | 0.397 | 0.73(1) | 0.486 |
| $\theta_D$(K) | 202 | 207(1) | 236 |
| $2\Delta/k_B T_c$ | - | 2.18(2) | - |
| Reported by: | Ref 12 | This work | Ref 12 |

**Table II.** Superconducting $T_c$ [defined as 90% of normal state resistivity; Fig. 2 (d)], charge

density wave $T_{CDW}$ transition temperatures [defined as peak in resistivity; Fig. 2 (b, c)] and fitting parameters of the CDW phase fluctuations scattering model for resistivity (see text).

Considerable change in local CDW fluctuation scattering for crystals without CDW strives to

increase conduction and is concomitant with the greatly enhanced superconducting $T_c$

values. The units for c, d and a are in $(10^{-8}\,\text{m}\Omega\text{cm/K}^5)$, $(10^{-5}\,\text{m}\Omega\text{cm/K}^2)$, $(10^{-4}\,\text{m}\Omega\text{cm/K})$, respectively. The units for b and $\rho_0$ are in $(\text{m}\Omega\text{cm})$. For crystals near the middle of the alloy series where CDW cannot be detected in resistivity the highest crystallographic disorder and consequently a substantial increase of $\rho_0$ [i.e. total $\rho(T)$ when $T \rightarrow 0$] are expected. However, the impurity-like scattering due to local CDW fluctuations (phase disorder scattering) term b becomes negative for that range of x, making the overall resistivity $\rho_0$ smaller or similar to x = 0 and x = 2 crystals (Table II, columns b and $\rho_0$.). L range denotes the temperature range of linear aT+b fit.

| $x$ | $T_c$(K) | $T_{CDW}$(K) | c | $c_{range}$ (K) | d | $d_{range}$ (K) | a | b | $L_{range}$ (K) | $\rho_0$ |
|---|---|---|---|---|---|---|---|---|---|---|
| 0 | 0.14 | 112(1) | 2.12(7) | 3-20 | 22.1(2) | 30-100 | 102.8(1) | 1.37(1) | 200-300 | 0.343 |
| 0.09(1) | 2.2(1) | 96(3) | 0.52(2) | 3-20 | 5.43(5) | 34-86 | 22.4(3) | 0.226(7) | 200-300 | 0.109 |
| 0.25(1) | 3.0(5) | 80(3) | 0.27(2) | 5-20 | 2.17(3) | 34-65 | 8.08(1) | 0.054(1) | 200-300 | 0.086 |
| 0.52(1) | 4.2(1) | – | 0.04(1) | 5-20 | – | | 1.6(1) | 0.0004(1) | 200-300 | 0.019 |
| 0.84(1) | 3.9(1) | – | 0.6(1) | 5-20 | – | | 17.4(1) | 0.0057(1) | 200-300 | 0.050 |
| 1.10(1) | 3.7(2) | – | 0.030(3) | 5-20 | – | | 2.64(1) | 0.0031(1) | 200-300 | 0.114 |
| 1.43(1) | 4.1(2) | – | 0.20(1) | 5-20 | – | | 6.32(1) | 0.0066(1) | 200-300 | 0.314 |
| 1.65(1) | 4.3(1) | – | 0.63(5) | 5-20 | – | | 29.4(1) | 0.043(1) | 200-300 | 0.427 |
| 1.98(1) | 4.0 (1) | 70(3) | 0.20(1) | 5-20 | 3.19(4) | 30-60 | 16.8(1) | 0.037(1) | 200-300 | 0.036 |
| 2.00(1) | 0.8 | 75(1) | 0.79(3) | 3-20 | 7.78(3) | 30-64 | 34.3(2) | 0.225(1) | 200-300 | 0.056 |

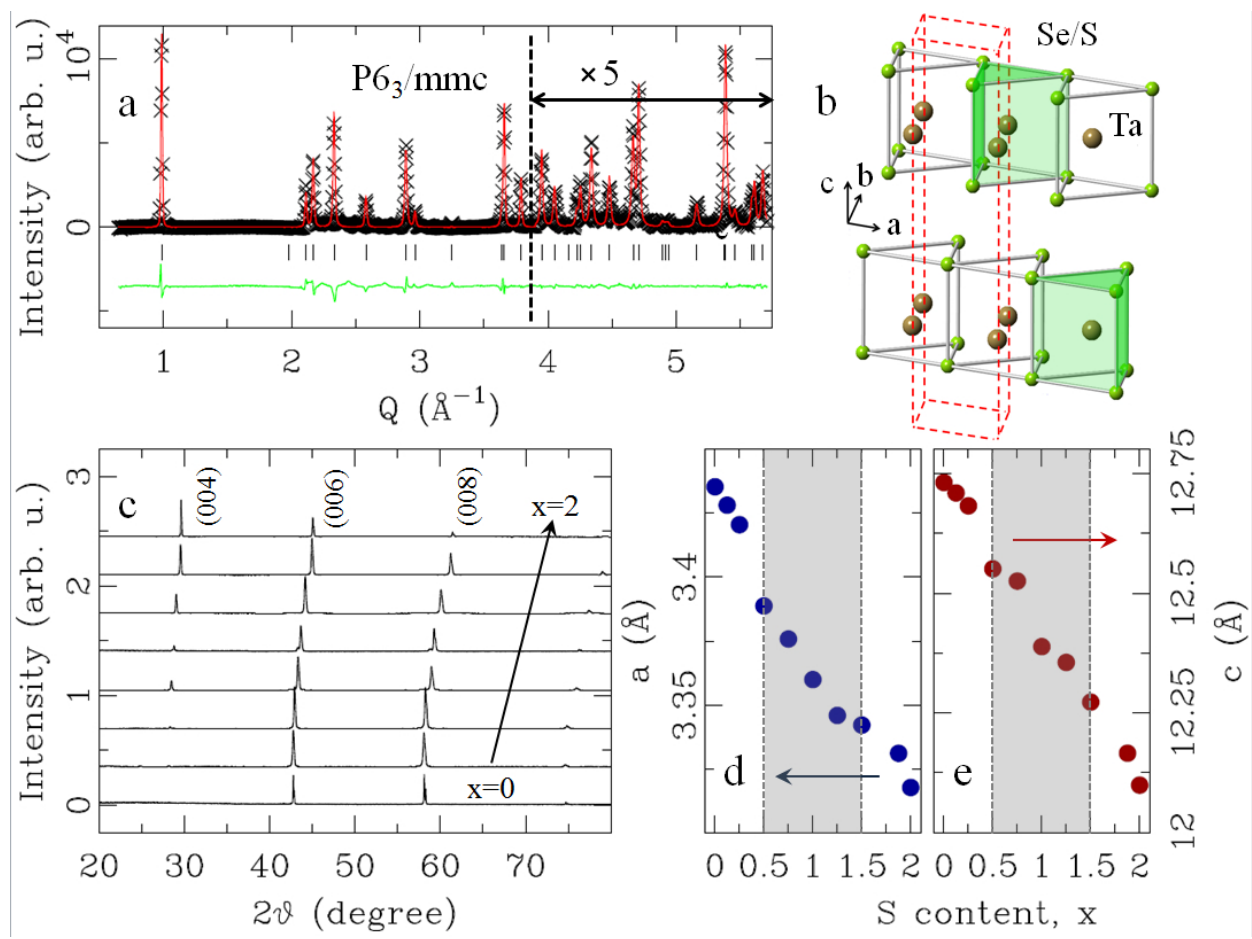

Figure 2

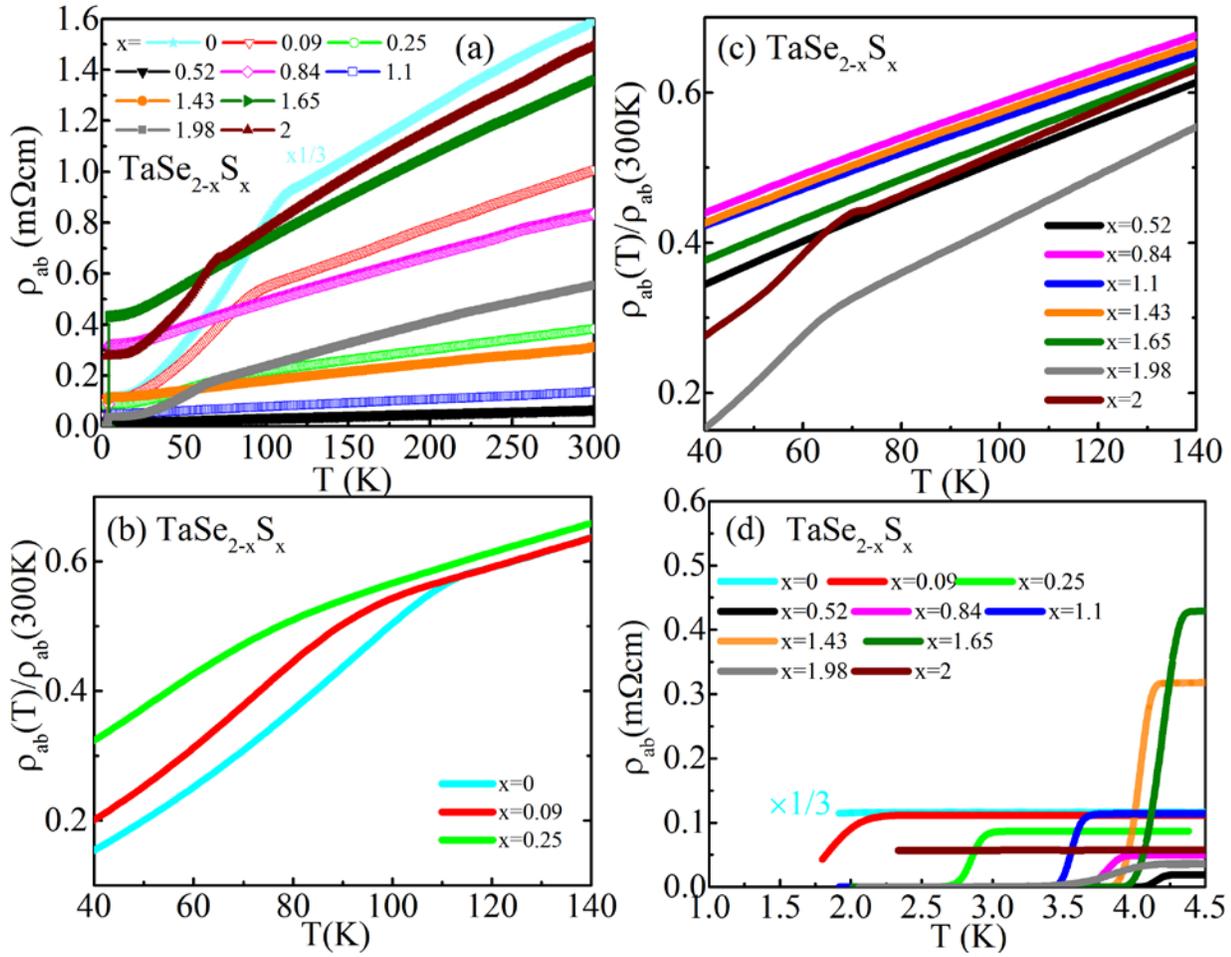

Figure 3

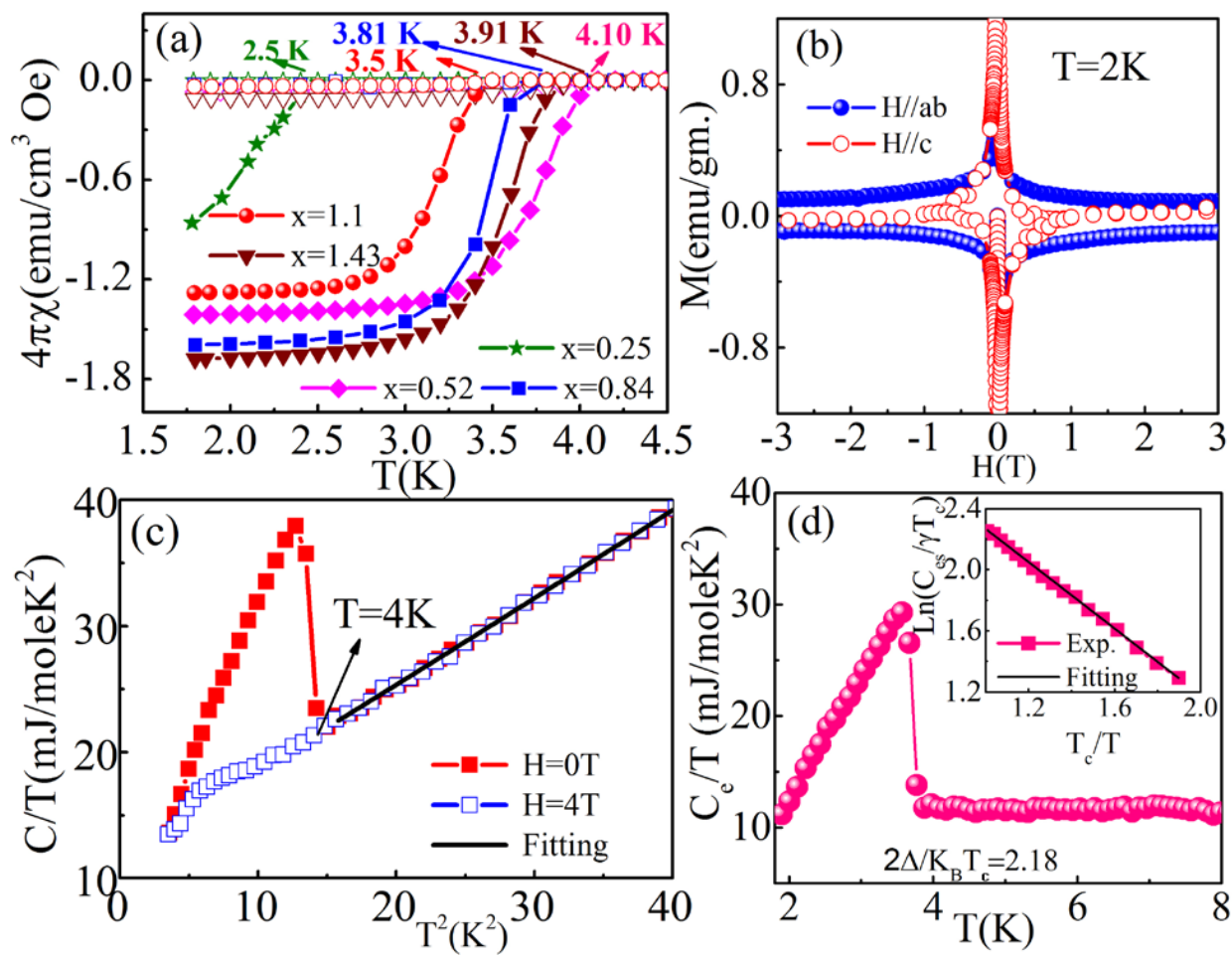

Figure 4

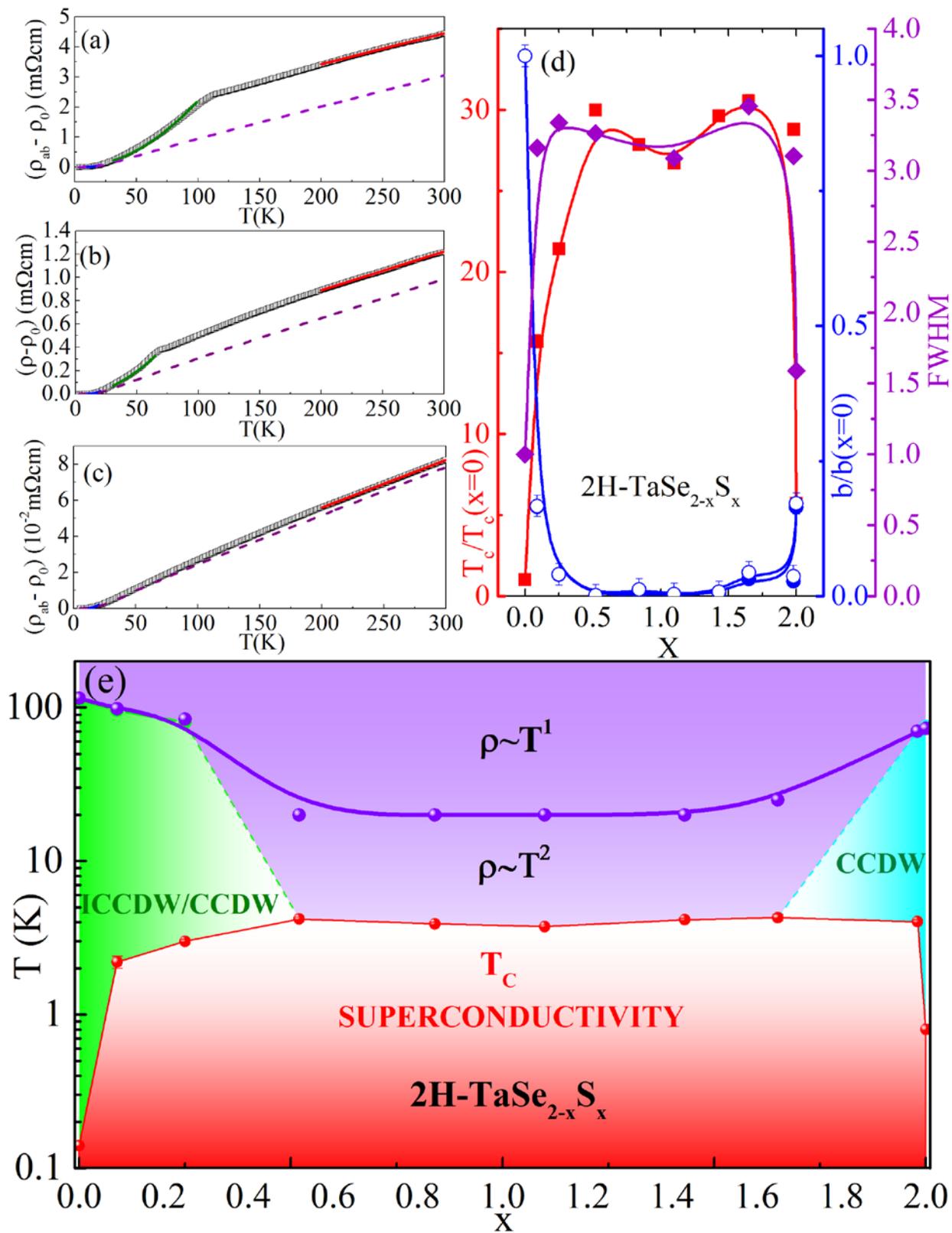

# Superconducting Order from Disorder in 2H-TaSe$_{2-x}$S$_x$

Lijun Li[1,2,†], Xiaoyu Deng[3], Zhen Wang[1], Yu Liu[1], A. M. Milinda Abeykoon[4], E. Dooryhee[4], A. Tomic[5], Yanan Huang[1,*], J. B. Warren[6], E. S. Bozin[1], S. J. L. Billinge[1,5], Y. P. Sun[2,7], Yimei Zhu[1], G. Kotliar[1,3] and C. Petrovic[1,†]

[1]*Condensed Matter Physics and Materials Science Department, Brookhaven National Laboratory, Upton, New York 11973, USA*
[2]*Key Laboratory of Materials Physics, Institute of Solid State Physics Chinese Academy of Sciences Hefei 230031 China*
[3]*Department of Physics & Astronomy, Rutgers, The State University of New Jersey - Piscataway, NJ 08854, USA*
[4]*Photon Sciences Directorate, Brookhaven National Laboratory, Upton NY 11973 USA*
[5]*Department of Applied Physics and Applied Mathematics, Columbia University, New York NY 10027 USA*
[6]*Instrumentation Division, Brookhaven National Laboratory, Upton, New York 11973, USA*[7]
[7]*High Magnetic Field Laboratory, Chinese Academy of Sciences, Hefei 230031 China*

†petrovic@bnl.gov and lilijun@issp.ac.cn; *Present address: Key Laboratory of Materials Physics, Institute of Solid State Physics, Chinese Academy of Sciences, Hefei 230031, China.

## Supplementary Information

### (A) Weak disorder in 2H-TaSe$_{2-x}$S$_x$ (0≤x≤2)

Here we justify the applicability of the Anderson theorem, stating that as long as $k_F l \gg 1$ (where $k_F$ is the Fermi wave vector and l is a mean free path), the $T_c$ is not affected by the presence of nonmagnetic impurities. We estimate the value of $k_F l$ in our samples as follows. In Boltzmann theory the 2D resistivity is: $\rho = (h/e^2)(d/k_F l)$ where h is Planck constant, e is elementary charge, $k_F$ is the Fermi wave number, d is interplane distance and l is mean free path. Taking approximately the value of lattice parameter c ∼ d ∼ 12.5 Å and ρ ∼ 0.1 mΩcm (Fig. 1 and Table II in our manuscript), we obtain $k_F l \sim 31$.

### (B) Electrical transport in 2H-TaSe$_{2-x}$S$_x$ (0≤x≤2)

As discussed in the main text, resistivity of 2H-TaSe$_{2-x}$S$_x$ ($0 \le x \le 2$) can be approximated [S1] as: $\rho(T) \sim aT + b$ above T$_{CDW}$ due to electron-phonon scattering and phase disorder impurity-like scattering off the local CDW fluctuations, $\rho(T) \sim dT^2$ immediately below T$_{CDW}$ due to scattering of electrons by collective excitations of CDW (the rapid drop just below CDW is due to CDW phase ordering) and $\rho(T) \sim cT^5$ at temperatures below about 15 K -20 K due to normal electron-phonon scattering. Attempts to extend linear fits in temperature range near T$_{CDW}$ resulted in unphysical negative values of resistivity (scattering), therefore we restrict linear fits to temperature above 200 K. As it can be seen [Fig. 4 (a-c) main text, Figure S1, Table II main text], resistivity of 2H-TaSe$_{2-x}$S$_x$ ($0 \leq x \leq 2$) can be well explained by the above arguments. Moreover, fit quality residuals in all temperature regions are satisfactory (Figure S2). Of our prime interest is the temperature independent linear term in high temperature resistivity aT+b.

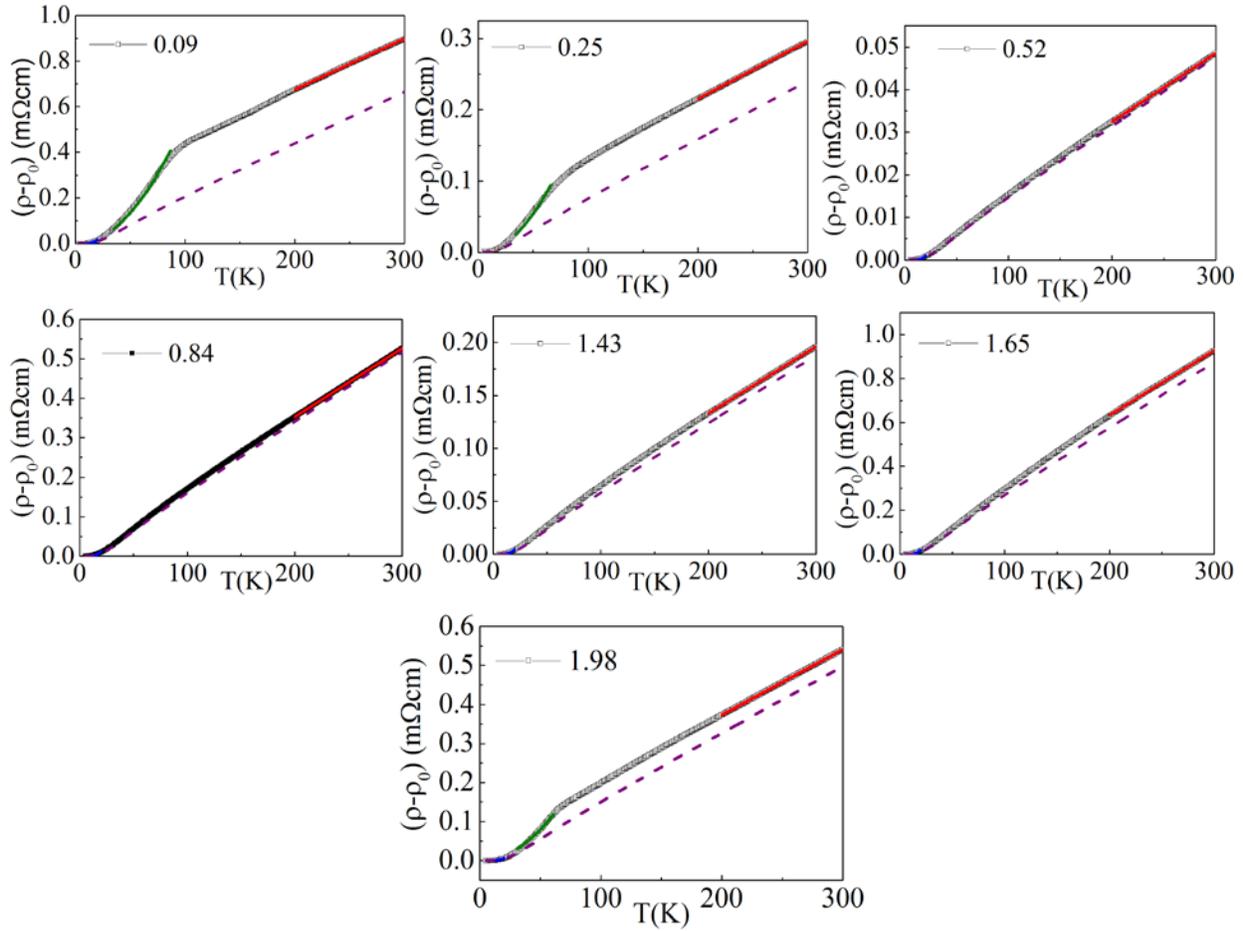

Figure S1: Electronic scattering mechanism for 2H-TaSe$_{2-x}$S$_x$ ($0 \le x \le 2$). Low temperature phonon scattering, scattering off collective excitations of CDW and high temperature scattering off the local CDW fluctuations are shown by blue, green and red solid lines respectively. Dashed violet line shows the phonon resistivity approximated by the Bloch-Grüneisen formula (see text in supplement).

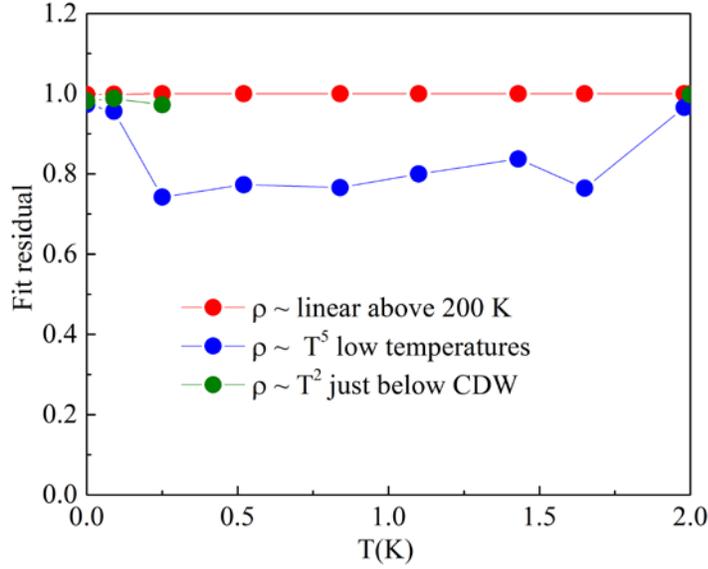

Figure S2: Fit quality for electronic scattering mechanism shown in Figure S1 (see text)

The above discussion excludes the contribution of phonon resistivity at high temperatures. In what follows we will show that in the asymptotic high temperature limit we observe the same trends for parameter b(x) in 2H-TaSe$_{2-x}$S$_x$ if we subtract the the Bloch-Grüneisen phonon resistivity. Phonon scattering is approximated by Block-Grüneisen formula [S1,S2]:

$\rho_P = 4\rho_\theta(T/\theta)^5 \cdot J_5(\theta/T)$ where $J_5(x) = \int_0^x \frac{z^5 dz}{(e^z - 1)(1 - e^{-z})}$, $\theta$ is the Debye temperature. For $T \geq \theta$ this becomes $\rho_P \sim \rho_\theta(T/\theta)$ and for $T \ll \theta$ we have $\rho_P \sim 497.6\rho_\theta(T/\theta)^5$ [S1]. Bloch-Grüneisen resistivity as a function of temperature is shown by the dashed purple line (Fig. S1) and the fit values of $\rho_\theta$ and $\theta$ are shown in Table S1. The subtraction of electron-phonon resistivity allows us to estimate the phase fluctuation scattering contribution following procedure in Ref. S1. The fits are shown in Fig. S3 and the values of fit parameters are shown in Table S1. Of interest is the high temperature linear term, specifically its temperature independent part

b. We show the values of b obtained from this procedure in Table S1 as $b_{BG}$ (BG denotes that phonon Bloch-Grüneisen resistivity has been subtracted). We see that $b_{BG}$ shows the same trend with x as parameter b from fits without phonon subtraction shown in Figure 4(a-c) main text and in Figure S1. The $b_{BG}$ and b are shown with blue solid and open symbols in Figure 4(d) in the main text.

**Table S1.** Parameter b (labeled as $b_{BG}$) from high temperature linear aT+b when phonon resistivity $\rho_P$ is subtracted. For comparison values of b from high T fits where $\rho_P$ is not subtracted (main text Table II) are also shown. The trend of change of $b_{BG}$ and b with x in 2H-TaSe$_{2-x}$S$_x$ is identical.

| $x$ | $\rho_0$ K) | θ(K) | b $_{BG}$ | b |
|---|---|---|---|---|
| 0 | 1.28(1) | 125(1) | 1.44(2) | 1.37(1) |
| 0.09(1) | 0.271(3) | 121(1) | 0.240(1) | 0.226(7) |
| 0.25(1) | 0.089(2) | 111(2) | 0.058(1) | 0.054(1) |
| 0.52(1) | 0.020(1) | 125(3) | 0.0015(1) | 0.0004(1) |
| 0.84(1) | 0.213(4) | 122(2) | 0.017(1) | 0.0057(1) |
| 1.10(1) | 0.032(1) | 120(2) | 0.0048(3) | 0.0031(1) |
| 1.43(1) | 0.076(1) | 120(2) | 0.0106(1) | 0.0066(1) |
| 1.65(1) | 0.364(6) | 124(2) | 0.063(1) | 0.043(1) |
| 1.98(1) | 0.239(2) | 142(1) | 0.052 (1) | 0.037(1) |
| 2.00(1) | 0.400(6) | 120(1) | 0.246(2) | 0.225(1) |

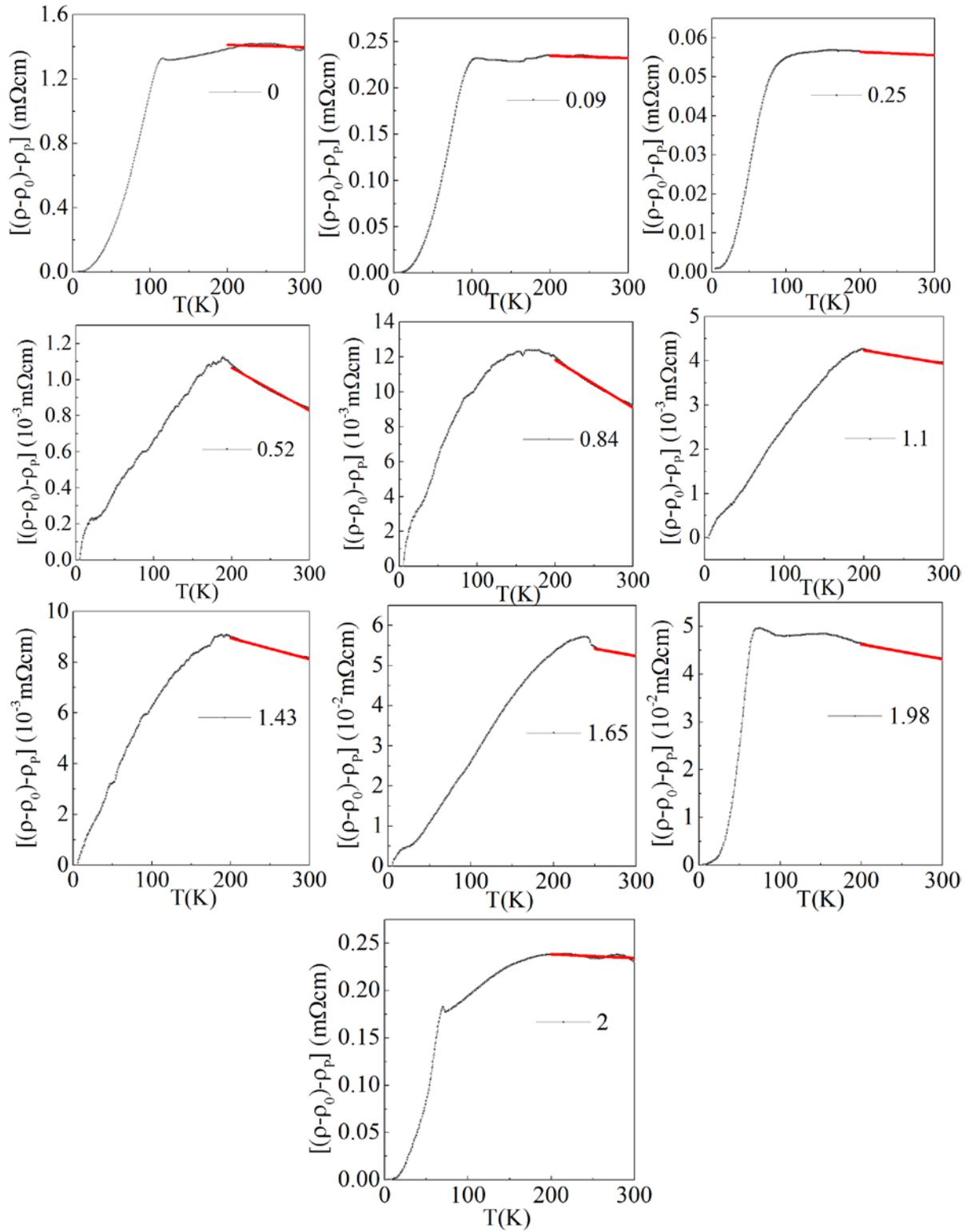

Figure S3: Estimate of the phase fluctuation scattering following the procedure in Ref. S1. The red line intercept at T=0 shows the parameter b from aT+b (see Fig. 5 in Ref. S1). The procedure gives nearly identical values with high temperature fits above 200 K for the raw data [Figure 4 in main text (a-d)]

From Table S1 and Fig. 4(d) in the main text (parameters b and $b_{BG}$ normalized to value of b and $b_{BG}$ for x=0) it is clear that the rapid increase of superconducting $T_c$ with x is concomitant with the decrease in the value of parameter b in the high temperature aT+b resistivity which arises from the local CDW fluctuations (S1). Smaller values of Debye temperature (θ) in Bloch-Grüneisen resistivity (Table S1) when compared to Debye temperature from heat capacity (Table I main text) is expected since only a fraction of phonons takes part in the electronic scattering.

## (C) Electronic structure calculation for 2H-TaS₂, 2H-TaSeS and 2H-TaS₂

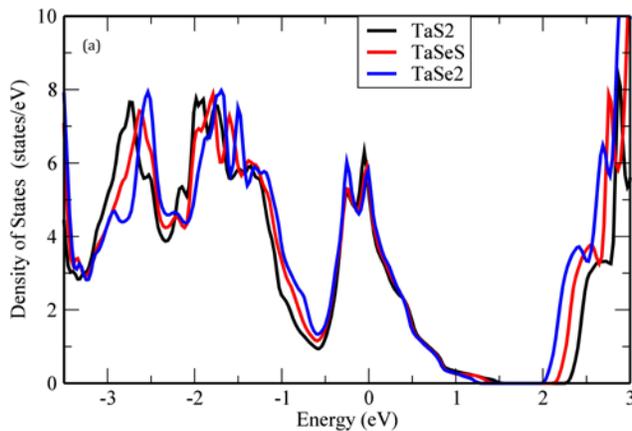

Figure S4: The LDA density of states of TaSe₂, TaS₂ and TaSeS

First principles calculations based on density functional theory were performed using the full-potential linearized augmented plane wave method, as implemented in the WIEN2k package [S3]. The local density approximation (LDA) of exchange-correlation functions was adopted and a 20×20×4 mesh was used to sample the Brillouin zone in the charge self-consistent calculation. Spin-orbital coupling effect is considered in the calculations. For TaSe₂ and TaS₂, we adopted their respectively experimental crystallographic structure. For TaSeS we assumed an artificial structure that is interpolated from the structures of TaSe₂ and TaS₂. And to mimic the doping, we assumed a layered commensurate substitution with two different patterns, S-S-Se-Se stacking and S-Se-S-Se stacking along c axis without site mixing and in-plane disorder. The results of these two situations are very similar so only the results of the former one are discussed below. We focus on the normal states without CDW order. Our results on TaSe$_2$ and TaS$_2$ are in accordance with previous calculations [S4-S6].

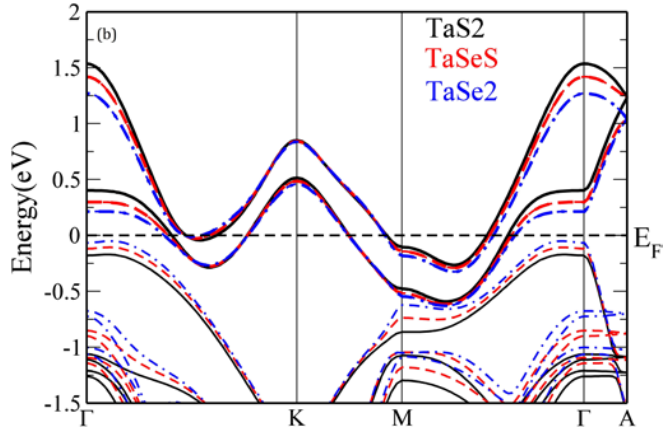

Figure S5: The band structures of TaSe$_2$, TaS$_2$ and TaSeS

Our results clearly demonstrate that the electronic structure of TaSe$_2$, TaS$_2$ and TaSeS are very similar in the normal states ( above the CDW transition temperatures of the end compounds) , as shown by their density of states [Figure S2(a) ] and band structures [Figure S2(b)]. Therefore the dramatic effects of doping, namely the enhancement of the superconducting transition temperature, are not the result of large changes in the hopping parameters. Instead we are dealing with emergent phenomena, where disorder suppresses the CDW ( a particle hole instability) leaving the s – wave superconductivity unaffected.

This picture is consistent with the specific heat coefficient derived from the values at the Fermi level from the LDA band structure. These are 12.26, 13.36, 12.96 mJ/mol K$^2$ for TaSe$_2$, TaS$_2$ and TaSeS respectively. The LDA predicted value of the specific heat coefficients are much larger than those measured ones of TaSe$_2$ (4.5 mJ/mol K$^2$ ) and TaS$_2$ (7.5 mJ/mol K$^2$) which are expected since the CDW order gaps part of the Fermi surface and thus reduces density of states at the Fermi level. It is instructive to note that the predicted specific heat coefficients are very close to the measured value (12.0 mJ/mol K$^2$ Table I in main text) in TaSe$_{1.48}$S0$_{.52}$. This is a strong indication that the density of states at the Fermi level are restored to that of the corresponding normal state, thus the CDW order is completely suppressed in the doped compounds. The fact that the measured values of the specific heat with the CDW suppressed in the doped compounds is close to the LDA value, suggests that strong correlations is not present in this material, and the resistitivity is due to scattering off local CDW fluctuations, an interpretation that was used in the main text.

**(D) TEM investigation of CDW**

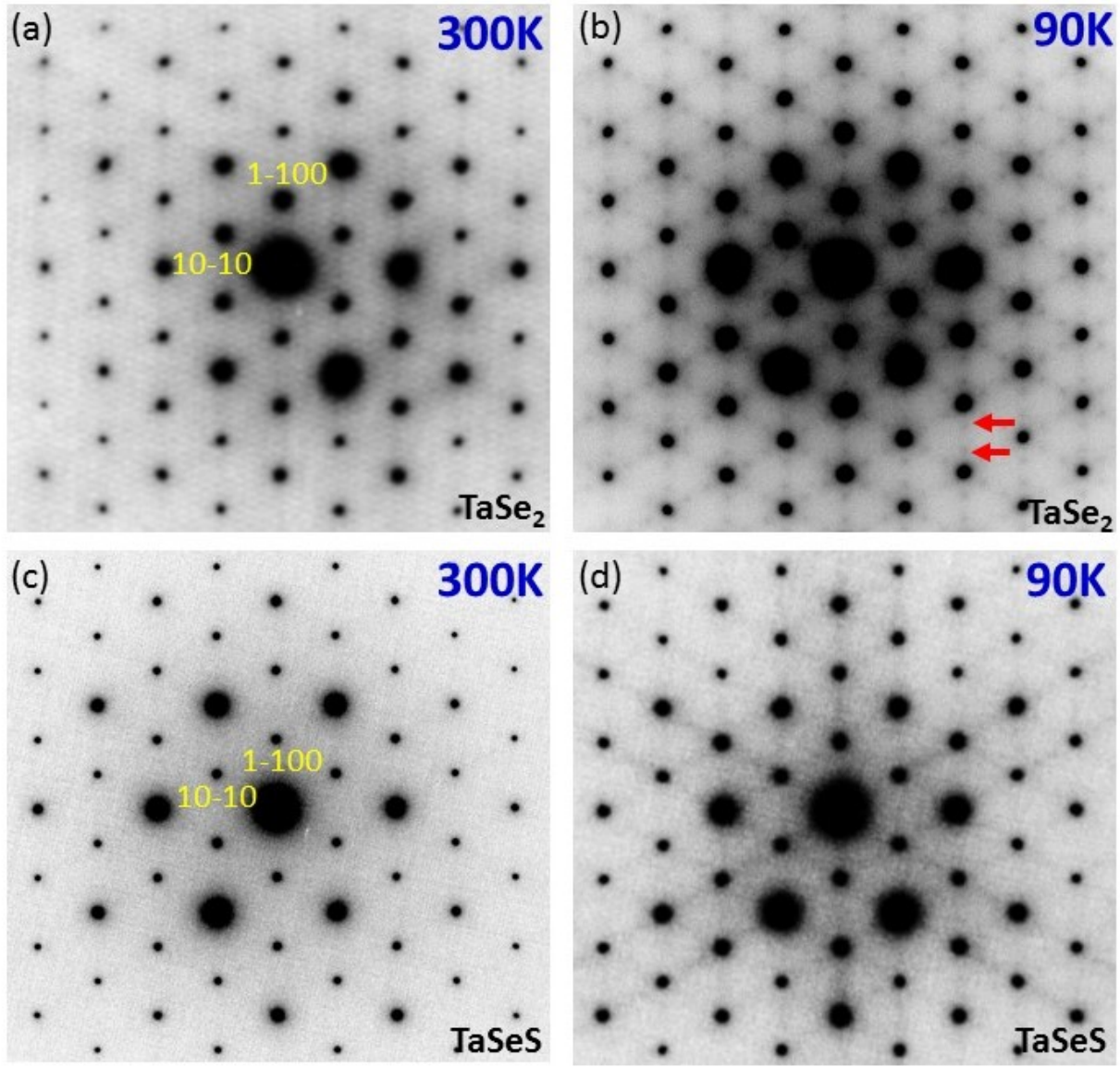

Figure S6: Diffraction patterns at 300 K and 90 K for 2H-TaS$_2$ and 2H-TaSe$_{0.9}$S$_{1.1}$.

In order to understand the microstructural feature of the CDW transition as shown in the R-T profile, we have performed experiments using transmitted electron microscopy(TEM) to study the structural/electronic changes with the decreasing of temperature. Selected area diffraction patterns (SADP) are shown in Figure S6 at room temperature(T=300K) and low temperature(T=90K) taken along [001] direction for 2H-TaSe$_2$ (a-b) and 2H-TaSe$_{0.9}$S$_{1.1}$ (c-d). The superlattice with modulation vector of q=1/3(1-100), as marked by the red arrows in (b), indicates the appearance of CDW in the pure 2H-TaSe$_2$. The appearance of CDW should correspond to the kink in the $\rho(T)$ . The absence (d) of uniform homogeneous supperlatice peaks in 2H-TaSe$_{0.9}$S$_{1.1}$ corresponds to the absence in kink in $\rho(T)$.

## (E) Hall Effect

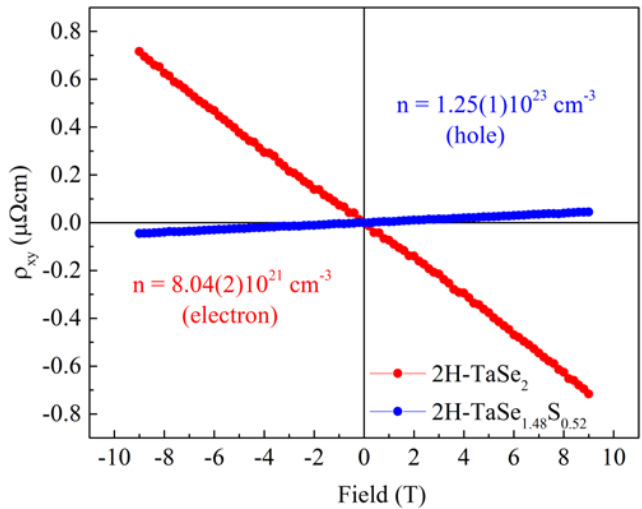

Figure S7: Hall effect measured at 50 K confirms the increase in carrier concentration with CDW suppression.

Hall effect measurements were performed in Quantum Design PPMS at T=50 K. Figure S7 shows data taken on 2H-TaSe$_2$ (T$_{CDW}$ = 112 K and superconducting T$_c$ = 0.14 K) and 2H-TaSe$_{1.48}$S$_{0.52}$ (T$_{CDW}$ = 0 and superconducting T$_c$ = 4.2 K). It can be seen that there is a significant increase in carrier concentration with doping and CDW suppression. In addition there is a change of the sign of dominant carrier from electron to hole.

## F) Elemental analysis from energy dispersive X ray spectroscopy (EDX)

Chemical composition of samples (Ta:Se:S ratio) is presented in Table II of the manuscript. There was no trace of iodine incorporation, as evidenced by EDX analysis of cleaved surfaces. Representative spectra is shown in Figure S8-S13.

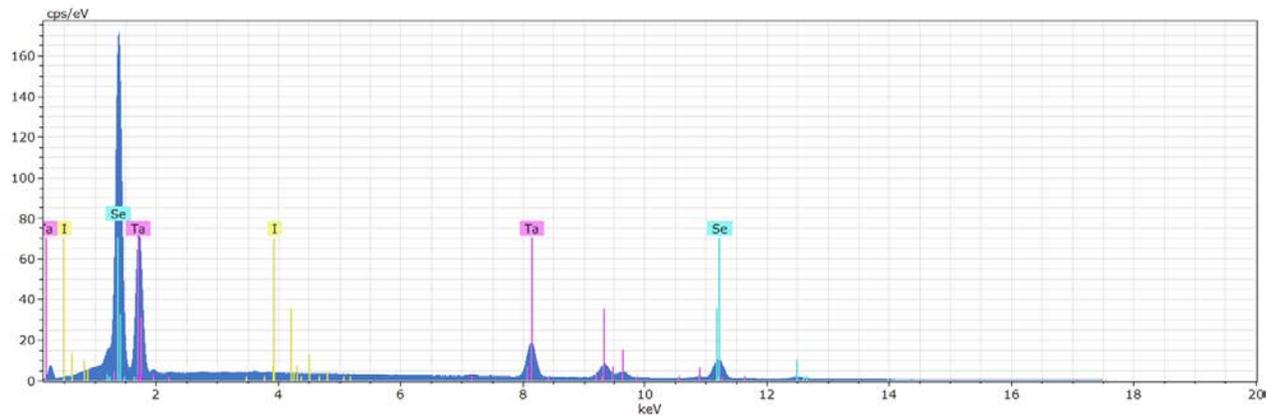

Figure S8: Elemental analysis of 2H-TaSe$_2$

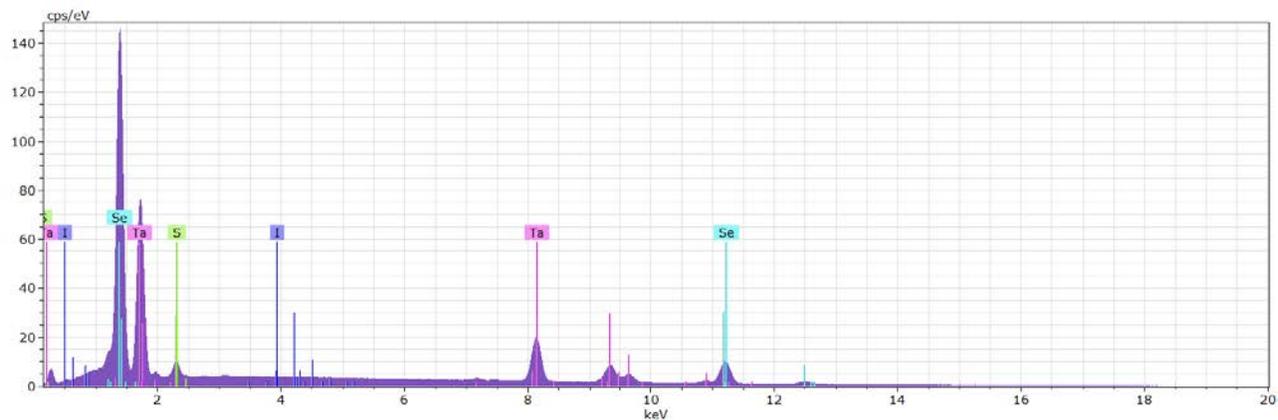

Figure S9: Elemental analysis of 2H-TaSe$_{1.75}$S$_{0.25}$

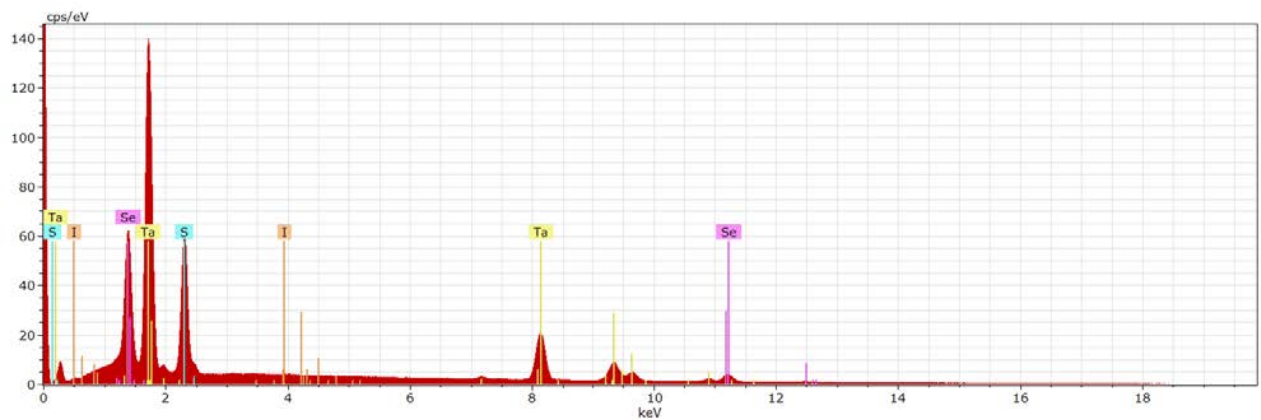

Figure S10: Elemental analysis of 2H-TaSe$_{1.48}$S$_{0.52}$

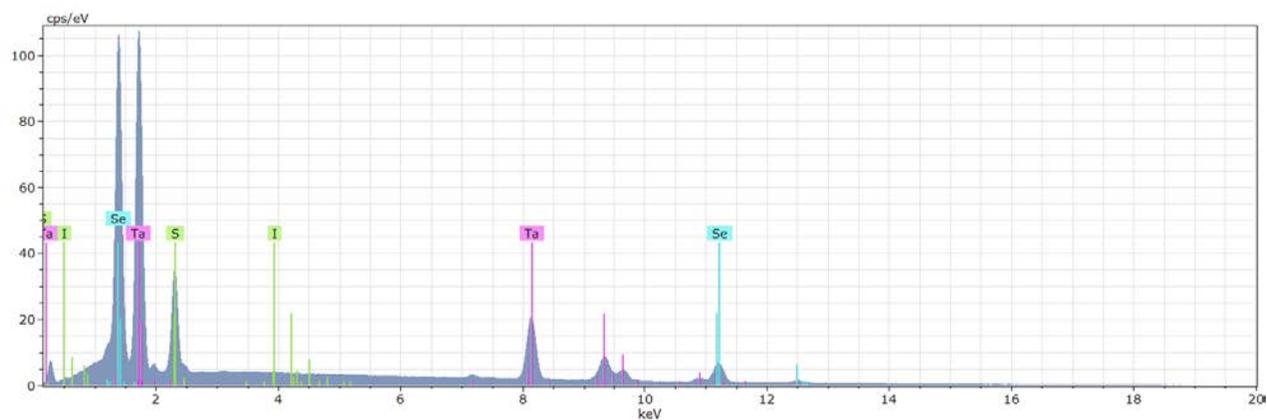

Figure S11: Elemental analysis of 2H-TaSe$_{0.9}$S$_{1.1}$

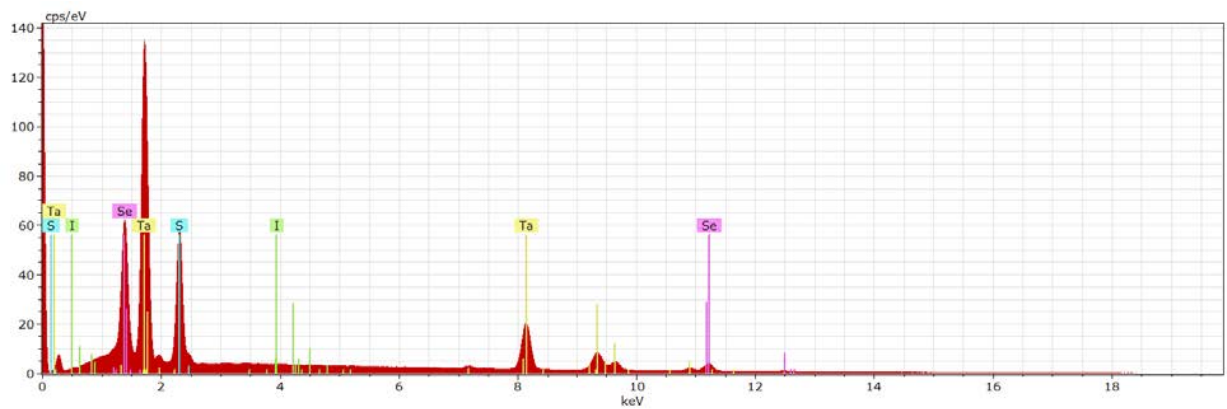

Figure S12: Elemental analysis of 2H-TaSe$_{0.35}$S$_{1.65}$

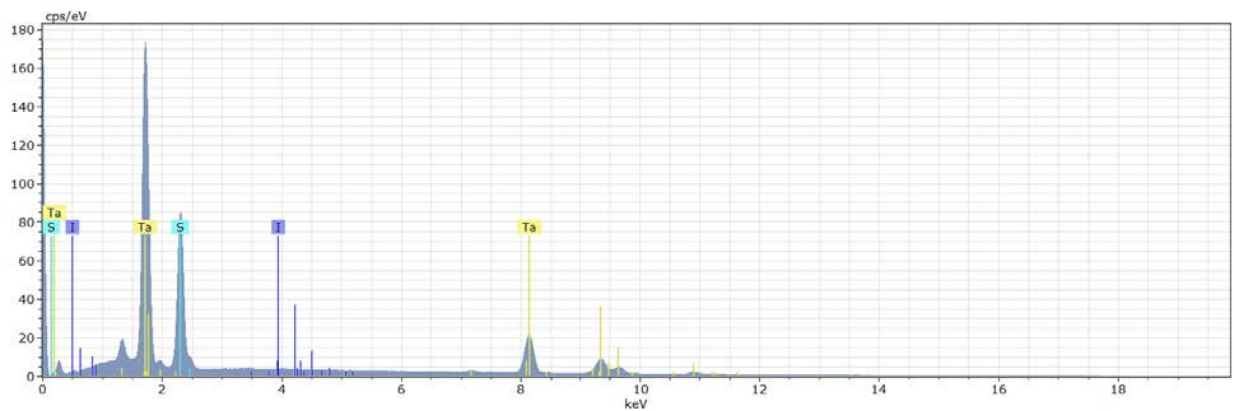

Figure S13: Elemental analysis of 2H-TaS$_2$

**Supplement References:**

[S1] M. Naito and S. Tanaka, Electrical transport properties in 2H-NbS$_2$, -NbSe$_2$, -TaSe$_2$ and -TaS$_2$, J. Phys. Soc. Jpn. 51, 219 (1982).

[S2] J. M. Ziman, *Electrons and Phonons* (Clarendon Press, Oxford, 1962).

[S3] P. Blaha, K. S., G. K. H. Madsen, D. Kvasnicka, and J. Luitz. WIEN2K, An Augmented Plane Wave + Local Orbitals Program for Calculating Crystal Properties,. edited by K. Schwarz, Technische Universitaet Wien, Austria (2001).

[S4] Yi Ding, Yanli Wang, Jun Ni, Lin Shi, Siqi Shi, Weihua Tang, First principles study of structural, vibrational and electronic properties of graphene-like MX2 (M=Mo, Nb, W, Ta; X=S, Se, Te) monolayers, Physica B: Condensed Matter, Volume 406, Issue 11, 15 May 2011, Pages 2254-2260

[S5] Jia-An Yan, Mack A. Dela Cruz, Brandon Cook and Kalman Varga, Structural, electronic and vibrational properties of few-layer 2H- and 1T-TaSe2, Scientific Reports 5, Article number: 16646 (2015)

[S6] J. Laverock, D. Newby, Jr., E. Abreu, R. Averitt, K. E. Smith, R. P. Singh, G. Balakrishnan, J. Adell, and T. Balasubramanian, k-resolved susceptibility function of 2H-TaSe2 from angle-resolved photoemission, Phys. Rev. B 88, 035108 (2013).